\documentclass[mode=notes, palette=ocean, fontsize=11pt]{modernclassnotes}

\usepackage{mathdots}

\course{Quantum Control Primer}
\coursecode{I. R. Sola}
\professor{Ignacio R Sola}
\institution{Department of Chemistry (SNU)}
\term{2014}
\docsubtitle{Lecture notes}
\title{Quantum Optimal Control Primer}

\begin{document}

\maketitle

\tableofcontents
\newpage

\lecture[]{1}{What Quantum Control Is About}

\section{The Quantum Control Problem}

These notes on Quantum Control provide the mathematical details of the lectures on Quantum Control delivered in the Department of Chemistry of Seoul National University. They focus only on the variational formulation of Quantum Control, also known as Quantum Optimal Control theory.
For other reviews that typically address the wider problem, the reader is refer to the general bibliography~\cite{Rice2000,RabitzSci00,Brumer2003,BrixnerCPC03,BalintKurtiACP08,EngelACP09,BrifACP12,HoffPCCP12,Ho2014,sola2018quantum}.

I will first refer to the more general Quantum Control (QC) problem.
I will later refine the definition of what QC is about, but to frame properly where
we are, let me just very generally refer to QC as the set of techniques that
allow you to manipulate a quantum system by means of external fields,
in order to drive the initial state to some predetermined outcome, in general
another quantum state more or less defined. 

In these notes the external
field will be a laser pulse. The range of parameters (frequencies, durations,
intensities, phases, modulations) that can be easily controlled in laser pulses,
and the coherent nature of the laser radiation, make laser pulses the best
tool to manipulate quantum systems. We will use strong fields, so that
the electromagnetic field will be treated classically. The use of quantum
fields poses interesting features that will not be addressed.

On the other hand, although our
presentation will be as general as possible, we will implicitly think about  
{\em isolated} molecules as our target system. Indeed, the isolated issue is
important and will be assumed throughout the notes, as well as the full
coherent nature of the interaction. Finally, in this work we will use the
dipole approximation (using the so called length gauge), so that the 
the time-depending Schr\"odinger equation of our system is

\begin{equation}
i \frac{d}{dt} | \psi \rangle = \left[ \hat{H}_0 - \vec{\mu} \vec{\epsilon}(t) 
\right] | \psi \rangle
\label{TDSE1}
\end{equation}

where $|\psi \rangle$ is the system's ket, $\hat{H}_0$ is the Hamiltonian 
operator of the isolated single molecule and $\vec{\mu}$ is the dipole operator 
that couples the laser field via the laser amplitude $\vec{{\epsilon}}(t)$ 
to the dynamics of the molecule. I am using and will use atomic units throughout these
notes, so $\hbar$ is not present in the equations.

Although the light field has vectorial properties (and there are interesting
effects depending on the polarization of the field) in this tutorial we will
be concerned with linearly polarized fields and we will often assume that
the molecule is perfectly oriented with respect to the field, such that
$\vec{\mu} \vec{\epsilon}(t) = | \mu | | \epsilon(t) | \cos(\theta) =
 \mu \epsilon(t)$, where we just remove the arrow from the vector and 
all quantities are scalar.

Eq.(1) represents conditions that can be met in the laboratory, particularly
in gas phase at low pressure using strong enough fields.
In this equation we have also disregarded all the spatial features of the laser.
 This can be done if the wavefront is essentially a plane wave as it hits a 
small, optically thin region, where the sample is located, which is valid
for optical (or larger) wave lengths and relatively small molecules.

The first step in solving the QC equations is necessarily being able to 
solve the time-dependent Schr\"odinger equation (TDSE) in the presence of a field. There 
are a number of techniques available and a few reviews where this material is
explained in detail~\cite{Kosloff1989,KosloffARPC94,tannor2007introduction}. We will assume that given the initial state $|\psi(0) \rangle$, 
the Hamiltonian and the field and a final time $T$, we know how to obtain 
$|\psi(T) \rangle$. Formally this can be written as

\begin{equation}
|\psi(T) \rangle = \hat{U}\left(T,0; \epsilon(t) \right) |\psi(0) \rangle
\end{equation}

where $\hat{U}\left( T,0; \epsilon(t)\right)$ is the time evolution 
operator, and we remark the parametric dependence with the field $\epsilon(t)$.
Notice that  $\hat{U}^{-1}\left(T,0; \epsilon(t)\right) = 
\hat{U}\left( 0,T; \epsilon(t) \right)$.
In reality one doesn't need to (and rarely does) solve the operator. 
What one does is to repeatedly map the action of a short time evolution 
operator on the wave function evaluated at the previous time

\begin{equation}
\begin{array}{l} 
\hspace{7cm} 
|\psi(\Delta t) \rangle = \hat{U}\left(\Delta t; \epsilon(0) \right) | \psi(0)\rangle \\
\hspace{4cm} 
|\psi(2\Delta t) \rangle= \hat{U}\left(\Delta t; \epsilon(\Delta t) \right) |\psi(\Delta t) \rangle \\ 
\hspace{5cm} \iddots   \\
|\psi(T) \rangle= \hat{U}\left(\Delta t; \epsilon(T-\Delta t)\right) |\psi(T-\Delta t) \rangle 
\end{array}
\end{equation}

In the limit $\Delta t \equiv dt \rightarrow 0$ the short-time evolution 
operator has a simple form
\begin{equation}
\hat{U}\left(dt; {\epsilon}(t) \right) = \exp {\left[ -i \left(
\hat{H}_0 - \mu {\epsilon}(t) \right) dt \right] }
\end{equation}

that can be approximated in different ways for finite $\Delta t$.
The overall time-evolution operator is the time-ordered product

\begin{equation}
\hat{U}\left(T,0; {\epsilon}(t)\right) = 
\hat{U}\left(\Delta t; {\epsilon}(T-\Delta t)\right) 
\cdots \hat{U}\left(\Delta t; {\epsilon}(\Delta t) \right) 
\hat{U}\left(\Delta t; {\epsilon}(0) \right) 
\end{equation}

but for many problems (mainly associated to grid-based representations 
of the wave functions) one rarely computes the matrix 
$\hat{U}\left(\Delta t; {\epsilon}(t) \right)$
but rather the vector $\hat{U}\left(\Delta t; {\epsilon}(t) \right) \psi(t)$,
which is computationally much advantageous.
 
{\em The essential problem in control} is the following: We know $\psi(0)$ and
we want our system in $\varphi_t$ at time $t = T$; so we must obtain the 
function ${\epsilon}(t)$ that brings $\psi(0)$ as closest as possible to 
$\varphi_t$. 

Theoretically there are three big questions related to the control problem

\begin{enumerate}

\item {\bf The what problem}. If indeed there exists a field ${\epsilon}(t)$ that brings $\psi(0)$ 
exactly (or differentially closed) to $\varphi_t$. In control terminology this is 
called the {\bf Controllability} problem~\cite{HuangJMP83,RamakrishnaPRA95,SchirmerPRA01,TuriniciJPA03}. 
Related to this problem is the {\em Quantum Connectivity} problem~\cite{WuPRA04}, which only looks at whether the target state overlaps with any unitary transformation of the initial state.

\item {\bf The how problem}. How to find an ``acceptable'' ${\epsilon}(t)$. This is normally 
achieved designing a variational problem where we get 
${\epsilon}(t) = {\epsilon}_{oc}(t)$ that maximizes the probability 
$$P_{0t} = |\langle \varphi_t | \hat{U}\left( T,0; {\epsilon}_{oc}\right) 
|\psi(0) \rangle |^2$$
This problem is called the {\bf Quantum Optimal Control} problem and the algorithms 
necessary to solve the problem are called quantum optimal control algorithms~\cite{PeircePRA88,Kosloff1989,SomloiCP93,ZhuJCP98,ZhuJCP98b,SolaJCPA98,SolaJPCA98b,OhtsukiJCP98,Sundermann1999a,TuriniciCP01,MadayJCP03,TuriniciJPA03,OhtsukiJCP04,WerschnikJOB05,WerschnikJPB07,OhtsukiPRA07,GollubPRL08,LapertPRA09,Castro2025}.

\item {\bf The why problem}. Which can be divided into two questions. The first one is what is the mechanism by which the control operates. The answer to this question allow us to {\em understand} and possibly {\em generalize} the type of solutions, whenever the system and the control can be classified under a certain class of families. One issue in understanding the {\bf mechanism} is that in Quantum Mechanics, a mechanism depends on the representation~\cite{MitraPRA03b,MitraPRA08}, which finally depends on the observer, which chooses which observables should one track (even if theoretically) to understand the evolution of the system.

The second question tries to map the global behavior of the system finding the properties of the space of all possible solutions, that is, 
how many different solutions there are and how sensitive the probability 
$P_{0t}$ is to changes in the optimal fields ${\epsilon}_{oc}(t)$,
what are the main control knobs in the field, what are the characteristics of
the dynamics, etc. These are questions very dependent on the Hamiltonian and
the optimal field. More generally one can try to find the global features of 
$P_{0t}$ for whatever initial and target state. This is called the
{\bf Landscape} or Optimal Landscape problem~\cite{RabitzScience2004,RabitzPRA2005,RabitzJCP2006,RabitzPRA06,ShenJCP2006,MoorePRA11}.

\end{enumerate}

In these notes we will only deal with the second problem of QC.
There are several ways to address it:

\begin{enumerate}

\item We call an {\bf open-loop design}\footnote{This nomenclature is not 
used for instance in the Engineering or Mathematical Control Theory or in Quantum 
Optics, but it is most often encounter in Molecular Physics. Perhaps more
appropriate would be to consider it as a constraint search, or few parameters
search approach} to an approximate maximization of  
$P_{0t} = |\langle \varphi_t | \hat{U}\left( T,0; {\epsilon}_{ap}\right) 
|\psi(0) \rangle |^2$
based on a motivated guess of a first order solution, $\epsilon_0(t)$
(chosen from analytic knowledge of the dynamics under a Hamiltonian 
$\hat{H}_{ap}(t)$ that looks similar to $\hat{H}(t)$ under certain limits)
whose parameters are fitted, typically by trial and error, until $P_{0t}$
is large enough.

\item We call {\bf local control} (LC) to the procedure by which one finds 
$\epsilon_{lc}(t)$ such that 
$$ \frac{d P_{0t}}{dt} = \frac{d}{dt} |\langle \varphi_t | 
\hat{U}\left( T,0; {\epsilon}_{lc}\right) |\psi(0) \rangle |^2 \ge 0
\hspace{1cm} \forall t$$

\item We call {\bf optimal control} (OCT) to the variational procedure by which one finds
$\epsilon_{oc}(t)$ such that $P_{0t}$ is a local maximum under variations of 
$\epsilon_{oc}(t)$, that is
$$ \frac{\delta P_{0t}}{\delta \epsilon_{oc}(t)} = 
\frac{\delta |\langle \varphi_t | \hat{U}\left( T,0; {\epsilon}_{oc}\right) |
\psi(0) \rangle |^2}
{\delta \epsilon_{oc}(t)} = 0 $$

\end{enumerate}

Below we will derive the equations that $\epsilon_{lc}(t)$ and $\epsilon_{oc}(t)$ must
meet. But before we get to this point, let me briefly mention other QC problems
related, but different, to the one that was stated above.
For instance one may ask for less detailed control problems
where instead of $\varphi_t$ the target is to maximize or minimize some 
expectation value, {\it e.g.} $\langle x \rangle$, $\langle H \rangle$, etc. 
at a certain time.
This involves less detailed target states and sometimes one can use simpler
methodologies, as {\bf local control theory}~\cite{SchwietersPRA91,BotinaJCP95,EngelACP09} or {\bf tracking}~\cite{ZhuJCP99,ZhuJCP03,MirrahimiJPCA05,SalomonJCP06,RothmanJCP05,RothmanPRA06}. 

Maximization of $P_{0t}$ can be rewritten as the maximization 
of the expectation value of the projector operator 
$\hat{P} =|\varphi_t \rangle \langle \varphi_t|$.
We will use this notation as the control formalism is then expressed in a more
general way, and one can easily move from the expectation value of $\hat{P}$
to the expectation value of another quantum operator $\hat{A}$.

On the other hand, one may ask for more detailed control problems where
instead of maximizing $P_{0t}$ one may want to maximize the
probability of getting different target states for any given initial state.
That is, we need to reconstruct an entire
unitary operator $\hat{U}_t$ with our field controlled propagator 
$\hat{U}\left( T,0; {\epsilon} \right)$, or in other words we
need to minimize the difference between $||\hat{U}_t - \hat{U}\left( T,0; 
 {\epsilon} \right) ||$, where the double $||$ indicates some
possible measure of distance or difference in the unitary matrices.
This is the key to using quantum control as a tool for 
Quantum Information or Quantum Computing problems~\cite{PalaoPRL02,TeschPRL02,PalaoPRA03,PalaoPRA08,Goerz_NJP2014}.
Finally, a different optimization problem, where the variational method is applied to the initial wave function, has been recently applied~\cite{ChangJPCL15,ChangJCTC15,ChangJPCA15,SampedroPCCP16,SampedroPCCP16b,ChangJPCA17,ArcosJPCL24,SolaCS24,SolaJCTC25,VeintemillasJCTC25}.

\section{Deriving the Quantum Optimal Control Equations}

We want to find the field that maximizes 
$J =\langle \psi(T) | \hat{P} | \psi(T) \rangle$, where $\hat{P}$ is a
projector operator, $\hat{P} = |\varphi_t \rangle \langle \varphi_t |$,
such that $J = P_{if}$ is the transition probability to go from $\psi(0) = \psi_i$ at
initial time to $\varphi_t$ at final time. The same derivation is valid
if $\hat{P}$ is any positive semi-definite operator (the jargon just means
that $\langle \psi(T) | \hat{P} | \psi(T) \rangle \ge 0$ for any $\psi$). 
At the same time
we want to penalize the use of very strong laser pulses, for two reasons:
(i) strong pulses break the validity of most approximate Hamiltonian 
descriptions, and (ii) we need an explicit dependence on the field in
our equations. As you will see below we need to include the field (at least)
up to power two in $J$.

To start up, we set-up a variational method to find the extremes of the 
functional

\begin{equation}
J =\langle \psi(T) | \hat{P} | \psi(T) \rangle - \frac{1}{T} 
\int \frac{\epsilon^2(t)}{f^2(t)} dt 
\end{equation}
We can vary $\psi(T)$ and $\epsilon(t)$.
Here $f(t)$ is a fixed function that makes the functional dimensional
free.

We can vary $\psi(T)$ and $\epsilon(t)$. Therefore we write

\begin{equation}
J + \delta J = \langle \psi(T) + \delta \psi(T) | \hat{P} | 
\psi(T) + \delta \psi(T) \rangle
- \frac{1}{T} \int \frac{\left[ \epsilon(t) + \delta \epsilon(t) \right]^2}
{f^2(t)} dt
\end{equation}

Taking into account only variations in first order of $\psi(T)$ and 
$\epsilon(t)$ (or equivalently, dropping all second order variations),
\begin{equation}
\delta J = \langle \delta \psi(T) | \hat{P} | \psi(T)  \rangle 
+ \langle \psi(T) | \hat{P} |  \delta \psi(T) \rangle
- \frac{2}{T} \int \frac{\epsilon(t)}{f^2(t)} \delta \epsilon(t) dt
\end{equation}

As the first two terms are complex conjugates, 
$z + z^* = 2 \mathrm{Re}(z)$, we can write
\begin{equation}
\delta J = 2\mathrm{Re} \left[ \langle \psi(T)| \hat{P}| \delta \psi(T) \rangle 
\right] - \frac{2}{T} \int \frac{\epsilon(t)}{f^2(t)} \delta \epsilon(t) dt
\label{dJ}
\end{equation}

In order to find a maximum (extreme) of $\delta J / \delta \epsilon(t)$ we
need to find how $\delta \psi(T)$ depends on $\delta \epsilon(t)$.
In most derivations of the quantum OCT equations one uses Lagrange multipliers 
to consider both functions independent~\cite{PeircePRA88,ShiJCP88,SolaJCPA98,WerschnikJPB07,Castro2025}. In this derivation we explicitly 
show how they relate to each other via the TDSE (the dynamical constraint).
The quantum dynamics of the system, including the variations, must follow
the equation

\begin{equation}
\dot{\psi}(t) + \dot{(\delta\psi)}(t) = -i \left[ \hat{H}(t) - 
\mu \delta\epsilon(t) \right] \left( \psi(t) + \delta\psi(t) \right)
\end{equation}

But remember that we only keep variations up to first order. Thus
\begin{equation}
\dot{(\delta\psi)}(t) = -i \hat{H}(t) \delta\psi(t)  + i\mu \delta\epsilon(t) 
\psi(t)
\end{equation}

This is a non-homogeneous TDSE that connects $\delta\psi(t)$ with 
$\delta\epsilon(t)$ at all times, and in particular, at final time,
as required in Eq.(\ref{dJ}).

The solution of a non-homogeneous first-order differential equation 
$\dot{y}(t) = ay(t) + b(t)$ is given by the sum of the solution of the 
homogeneous equation plus the particular solution:
\begin{equation}
y(t) = e^{at} y(0) + \int_0^t e^{a(t-t')} b(t') dt'
\end{equation}
as you can check by inspection (derive the solution).

Applied to our TDSE we find
\begin{equation}
\delta\psi(t) = \hat{U}(t,0;\epsilon) \delta\psi(0) + i \int_0^t 
\hat{U}(t,t';\epsilon(t')) \mu \delta\epsilon(t') \psi(t') dt'
\end{equation}
as you can also check by inspection.

We are not allowed to change the initial state. Thus, $\delta\psi(0) = 0$.
At final time 
\begin{equation}
\delta\psi(T) =   i \int_0^T \hat{U}(T,t';\epsilon(t')) 
 \mu \delta\epsilon(t') \psi(t') dt'  \label{dpsi}
\end{equation}

Substituting Eq.(\ref{dpsi}) in Eq.(\ref{dJ})  we obtain

\begin{equation}
\delta J = \int_0^T dt' \left\{ 2\mathrm{Re}\left[ i \langle \psi(T) |
\hat{P} \hat{U}(T,t';\epsilon) \mu \delta\epsilon(t') | \psi(t') \rangle 
\right] - \frac{2}{Tf^2(t')} \epsilon(t') \delta \epsilon(t') \right\}
\end{equation}

Since $\delta \epsilon(t)$ is real we can take it out of the real part.
Additionally, given a complex function $z$, $\mathrm{Re}(iz) = 
-\mathrm{Im}(z)$ and thus
\begin{equation}
\delta J = -2 \int_0^T dt' \left\{ \mathrm{Im} 
\left[ \langle \psi(T) | \hat{P} \hat{U}(T,t';\epsilon) \mu |\psi(t')\rangle 
\right] + \frac{1}{Tf^2(t')} \epsilon(t') \right\}  \delta \epsilon(t')
\end{equation}

We can define an instantaneous gradient $d J_t / d\epsilon(t)$ such that
\begin{equation}
\delta J = \int_0^T \frac{d J_t}{d\epsilon(t)} \delta \epsilon(t) dt
\end{equation}
where
\begin{equation}
\frac{d J_t}{d\epsilon(t)} = -2 \left\{ \mathrm{Im} 
\left[ \langle \psi(T) | \hat{P} \hat{U}(T,t;\epsilon) \mu |\psi(t)\rangle 
\right] + \frac{1}{Tf^2(t)} \epsilon(t) \right\}  \label{grad1}
\end{equation}

Next we find a maximum of the functional making 
$d J_t / d\epsilon(t) = 0$ at each time $t$.
This equation is obeyed by the optimal field. We get

\begin{equation}
\mathrm{Im} \left[ \langle \psi(T) | \hat{P} \hat{U} \left(T,t;\epsilon_{oc}(t)
\right) \mu | \psi(t) \rangle \right] + \frac{1}{T f^2(t)} \epsilon_{oc}(t) = 0
\end{equation}

Which gives us
\begin{equation}
\epsilon_{oc}(t) = - T f^2(t) \mathrm{Im} \left[ \langle \psi(T) | 
\hat{P} \hat{U}\left(T,t;\epsilon_{oc}(t)\right) \mu | \psi(t) \rangle \right]
\end{equation}
or
\begin{equation}
\epsilon_{oc}(t) = - T f^2(t) \mathrm{Im} \left[ \langle \psi(T) | 
\hat{P} \hat{U}\left(T,t;\epsilon_{oc}(t)\right) \mu 
\hat{U}\left(t,0;\epsilon_{oc}(t)\right)| \psi(0) \rangle \right]
\label{Eop2}
\end{equation}

This is an implicit equation: $\epsilon_{oc}(t)$ depends on 
$\hat{U}\left(T,t;\epsilon_{oc}(t)\right)$ which itself depends on
$\epsilon_{oc}(t)$ in a complicated way. In particular, knowing 
$\epsilon_{oc}(t)$ requires knowledge of the whole history of the dynamics,
from the past to the present [via $\hat{U}\left(t,0;\epsilon_{oc}(t)\right)$]
and from the future to the present 
[via $\hat{U}\left(T,t;\epsilon_{oc}(t)\right)$].
To solve the non-linear equation one needs iterative procedures.
Notice that, had we not included the quadratic penalty form of the field
in the functional, the gradient [Eq.(\ref{grad1})] wouldn't be an 
explicit function of the field. Then we couldn't have obtained Eq.(\ref{Eop2}).

Usually we write Eq.(\ref{Eop2}) in more symmetrical fashion.
First we define the performance function, 
\begin{equation}
\chi(T) = \hat{P} \psi(T) 
\end{equation}
Then, notice that
\begin{equation}
\langle \psi(T) | \hat{P} \hat{U}(T,t;\epsilon_{oc}) = \langle \chi(T) | 
\hat{U}(T,t;\epsilon_{oc}) = \langle \chi(t) |  \label{chit}
\end{equation}
The performance function
can only be evaluated at the end of the dynamics (it's fixed
at time $t = T$), and then we propagate it backwards in time with the
optimal field, which we can write as

\begin{equation}
\epsilon_{oc}(t) = - T f^2(t) \mathrm{Im} \left[ \langle \chi(t) | 
 \mu | \psi(t) \rangle \right]  \label{Eop}
\end{equation}

This is the Quantum OCT equation for the optimal field.
It is very compact and deceivingly simple, but do not forget that both
$\chi(t)$ and $\psi(t)$ depend on the field, so the equation (obviously)
remains nonlinear. Notice that everything is fine with the dimensions:
$f(t)$ has the dimensions of a field,
$\mu \times f(t) \times T$ has the dimensions of $\hbar$, which is
in the denominator (it is absent in our equation as we use atomic units).
You are free to define a positive envelope function $S(t) \equiv f^2(t)$
and to use any scale factor $\alpha$ instead of $T$ in the equation.
Then you get the equation most often found in the publications:

\begin{equation}
\epsilon_{oc}(t) = - \frac{\alpha}{\hbar} S(t) \mathrm{Im} 
\left[ \langle \chi(t) | \mu | \psi(t) \rangle \right]  \label{Eopb}
\end{equation}

We can also write the gradient [Eq.(\ref{grad1})] in terms of the performance 
function
\begin{equation}
\frac{d J_t}{d\epsilon(t)} = -2 \left\{ \mathrm{Im} 
\left[ \langle \chi(t) | \mu |\psi(t)\rangle 
\right] + \frac{1}{Tf^2(t)} \epsilon(t) \right\}  \label{grad2}
\end{equation}

\section{Finding the Optimal Control Pulses}

As noticed, Eq.(\ref{Eop}) [or the equivalent forms (\ref{Eop2}), (\ref{Eopb})]
are implicit equations that can only be solved by iterative procedures.
Here we are not going to derive the most common algorithms that solve the 
optimal control equations nor study their convergence analysis. 
Rather I am just going to comment the two that I mostly use.
Since they imply iterative procedures, I will use
$\epsilon^{(k)}(t)$ and $\epsilon^{(k+1)}(t)$ as two consecutive
steps in the long run to obtain good (optimal) solutions. To start the algorithm
running you always need an initial guess $k=0$. In some occasions a very
nonimaginative guess $\epsilon^{(0)}(t) = 0$ is sufficient to end-up
with good solutions. On the other hand you also need to give up at a
certain point. Normally this is achieved by imposing convergence conditions,
for instance 
$$  \int_0^T dt \frac{\left[ \epsilon^{(k+1)}(t) - 
\epsilon^{(k)}(t) \right]^2}{f(t)^2} \le T\kappa \hspace{1cm} \mathrm{or}
\hspace{1cm} J^{(k+1)} - J^{(k)} \le \kappa'$$
where $\kappa$ ($\kappa'$) are predetermined criteria for convergence.

First, let me tell you what does not work. Unfortunately, from Eq.(\ref{Eop2})
the obvious fixed-point iteration:
$$ \epsilon^{(k+1)}(t) = - T f^2(t) \mathrm{Im} \left[ \langle \psi(T) | 
\hat{P} \hat{U}\left(T,t;\epsilon^{(k)}(t)\right) \mu 
 \hat{U}\left(t,0;\epsilon^{(k)}(t)\right)| \psi(0) \rangle \right]$$
does not work in general.

To simplify the notation and use Eq.(\ref{Eop}) I will call 
$\psi^{(k)}(t)$ to the wave function obtained by propagating with 
$\hat{U}\left(t,0;\epsilon^{(k)}(t)\right)$ and likewise
$\chi^{(k)}(t)$ to the performance function obtained by 
propagating with $\hat{U}\left(T,t;\epsilon^{(k)}(t)\right)$.

In a following section we will study the algorithms in detail.
We here refer to two families of methods that depend on evaluating the gradient, either to use it for a line-search along the direction of the gradient or conjugate-gradient (what we call the gradient method) or a generalized version of the gradient that allows a ``large-step'' search (what we call a Krotov-type algorithm).
Notice that there are other (slower) methods, that do not require evaluating the derivative, that may use learning algorithms~\cite{JudsonPRL92}, simplex methods~\cite{TuriniciPRE04,Sola_PRA2023,Sola_AIPadv2023} or the CRAB method~\cite{CanevaPRA11,DoriaPRL11}.

\begin{enumerate}

\item The {\bf gradient method}, which is based on choosing the field on
step $k+1$ as the field on step $k$ plus a contribution along the gradient
obtained with the same $k$ field
\begin{equation}
\frac{d J_t^{(k)}}{d\epsilon(t)} = -2 \left\{ \mathrm{Im} 
\left[ \langle \chi^{(k)}(t) | \mu |\psi^{(k)}(t)\rangle 
\right] + \frac{1}{Tf^2(t)} \epsilon(t) \right\}
\end{equation}
Then 
\begin{equation}
\epsilon^{(k+1)}(t) = \epsilon^{(k)}(t) + \beta \frac{d J_t^{(k)}}{d\epsilon(t)}
\end{equation}
Where $\beta$ (a small but unknown positive number) must be found by linear 
search.
It is not difficult to prove that by this choice $J^{(k+1)} \ge J^{(k)}$
(at least for a certain range of $\beta$).

\item The {\bf Krotov method} is based on using as the optimal field at
step $k+1$
\begin{equation}
\epsilon^{(k+1)}(t) = - T f^2(t) \mathrm{Im} \left[ \langle \chi^{(k)}(t) | 
 \mu | \psi^{((k+1)}(t) \rangle \right]  \label{EopK}
\end{equation}

This equation is very similar to Eq.(\ref{Eop}), but notice that while
$\chi^{(k)}(t)$ is propagated using the field obtained at the previous step
(the only one that is known at all times), $\psi^{(k+1)}(t)$ is propagated
with the new field. This implies immediate feedback. Obviously some
numerical approximation must be done at some point, since we do not know
$\epsilon^{(k+1)}(t)$. However, numerically we can force $\epsilon^{(k+1)}(0)
= \epsilon^{(k)}(0)$ and then use this value, together with $\chi^{(k)}(t)$
to calculate via Eq.(\ref{EopK}) $\epsilon^{(k+1)}(\Delta t)$ and so on.
The immediate feedback will work better the smaller $\Delta t$ is.

It can also be proved that the Krotov method guarantees $J^{(k+1)} \ge J^{(k)}$
and {\em we do not need any line search!}

\end{enumerate}

\section{Deriving the Local Control Equations}

At the introduction I told you that there are several quantum control
procedures to manipulate the dynamics. I talked about the open-loop 
aproach, the optimal control approach and the local control approach.
Here we will focus on the later.

Following the original discussion, in LC we want to find the field 
$\epsilon_{lc}(t)$ that satisfies at all times

\begin{equation}
\frac{d P_{0t}}{dt} = \frac{d}{dt} 
|\langle \varphi_t | \hat{U}\left( T,0; {\epsilon}_{lc}(t) \right) |\psi(0) 
\rangle |^2 = \frac{d}{dt}  \langle \psi(t) | \hat{P} | \psi(t) \rangle \ge 0
\end{equation}

In fact, since in LC the operator $\hat{P}$ does not need to be positive semi-definite,
we will generalize the discussion so that LC is any procedure that seeks a 
monotonic increase or decrease in time of any observable of interest in the 
dynamics, be it a transfer probability $P_{0t}$ or a property like say, the 
energy of the system $\langle H_0 \rangle$.
Therefore we shall find $\epsilon_{lc}(t)$ that satisfies

\begin{equation}
\frac{d}{dt}  \langle \psi(t) | \hat{A} | \psi(t) \rangle \gtrless 0
\end{equation}

The time derivative of the expectation value follows from the time derivative of
the ket, $| \dot{\psi} \rangle = -i \hat{H} | \psi \rangle$  and that of the bra
$\langle \dot{\psi} | = i \langle \psi | \hat{H}$ (which is just the complex conjugate).
Thus

\begin{equation}
\frac{d}{dt}  \langle \psi(t) | \hat{A} | \psi(t) \rangle = i 
\langle \psi(t) | \hat{H} \hat{A} | \psi(t) \rangle - i \langle \psi(t) | \hat{A} \hat{H} 
| \psi(t) \rangle  \label{LC1}
\end{equation}
which can be written in terms of the commutator $\left[ \hat{A}, \hat{H} \right] =
\hat{A} \hat{H} -  \hat{H}\hat{A}$
\begin{equation}
\frac{d}{dt}  \langle \psi(t) | \hat{A} | \psi(t) \rangle = -i 
\langle \psi(t) | \left[ \hat{A}, \hat{H} \right] | \psi(t) \rangle \label{EV}
\end{equation}

It proves useful to write Eq.(\ref{LC1}) in the equivalent way
\begin{equation}
\frac{d}{dt}  \langle \psi(t) | \hat{A} | \psi(t) \rangle =
-2 \mathrm{Re} \left( i \langle \psi(t) | \hat{A}\hat{H} | \psi(t) \rangle \right)
= 2 \mathrm{Im} \left( \langle \psi(t) | \hat{A}\hat{H} | \psi(t) \rangle \right)
\end{equation}
since this reminds us that the expectation value is a real number despite
the presence of the imaginary unit (it is there only during the calculation,
not at the final result).
Likewise, the local control field can be chosen strictly real although
some "$i$" shows up in the equations.

Since $\hat{H}(t) = \hat{H}_0 - \mu \epsilon(t)$, separating the contributions
of the two parts of the Hamiltonian in Eq.(\ref{EV}) we obtain
\begin{equation}
\frac{d}{dt}  \langle \psi(t) | \hat{A} | \psi(t) \rangle =
-i \langle \psi(t) | \left[ \hat{A}, \hat{H}_0 \right] | \psi(t) \rangle
+i \epsilon(t) \langle \psi(t) | \left[ \hat{A}, \mu \right] | \psi(t)\rangle 
 = h(t) + \epsilon(t) g(t)
 \label{EV2}
\end{equation}
where both $h(t)$ and $g(t)$, as expectation values, are real functions of time.
Clearly, if $\left[ \hat{A}, \mu \right] = 0$ no control is possible 
(or at least we will not obtain any explicit dependence of our objective 
with the local control field). On the other hand, if $\left[ \hat{A}, 
\hat{H}_0 \right] = 0$ the control is simple,
since then, from Eq.~(\ref{EV2}) it is easy to impose 
$d\langle A(t) \rangle / dt \ge 0$ (or $\le0$). 
It suffices to make $\epsilon(t) g(t) \ge 0$, or
\begin{equation}
\epsilon_{lc}(t) = f(t) g(t) = i f(t) \langle \psi(t) | \left[ \hat{A}, \mu 
\right] | \psi(t)\rangle
= -2 f(t) \mathrm{Im} \left( \langle \psi(t) | \hat{A} \mu | \psi(t) \rangle \right)
\label{Elc}
\end{equation}
with $f(t) \ge 0$, since then $d\langle A(t) \rangle / dt = f(t) g(t)^2$.
(For $d\langle A(t) \rangle/dt \le 0$ we only need to choose
$f(t) \le 0$.) At the times when $g(t) = \left[ \hat{A}, \hat{\mu} 
\right] = 0$ (or the imaginary part of 
$\langle \psi(t) | \hat{A} \mu | \psi(t) \rangle$ is zero) 
the rate of change will be zero as well, provided
$h(t) = 0$. If the last condition is not met, there is no easy way
to know if that term will not cause a drop in
$\langle A(t) \rangle$. This is why in general, local control procedures
are only defined for operators $\hat{A}$ such that  
$\left[ \hat{A}, \hat{H}_0 \right] = 0$.

Local control is a {\em local-in-time} procedure. We need to
know the whole history of the wave function evolution in order to decide
what is the optimal field. To find a LC pulse we just need to adjust
the pulse instantaneously to the value of the wave function.
Inspecting Eq.(\ref{Elc}) more closely (I just choose one of the two
expressions, it could be any):
\begin{equation}
\epsilon_{lc}(t) = -2 f(t) \mathrm{Im} \left( \langle \psi(0) | 
\hat{U}\left( 0, t; \epsilon_{lc}(t) \right) \hat{A} \mu 
\hat{U}\left( t, 0; \epsilon_{lc}(t) \right) | \psi(0) \rangle \right)
\end{equation}
we find that we need to know $\epsilon_{lc}(t)$ in order to get
$ \epsilon_{lc}(t)$. This is the {\em instantaneousness} that I was
referring before. But any (numerical) discretization procedure aches
to getting $\psi(t+\Delta t)$ from $\psi(t)$ and $\epsilon(t)$.
Just fixing $\epsilon_{lc}(0)$ is enough to carry the integrations
and get $\epsilon_{lc}(t)$.
You do not need anything else than to be careful with respect to how
large $\Delta t$ is (because LC requires a fast adjustment of the field to
the wave function) in order to get sensible results from the local control 
equations. No feedback (future-to-present) integration is required.

Sometimes one can be overambitious and try to force $\langle \psi(t) | \hat{A} |
\psi(t) \rangle$ to follow a given trajectory $a(t)$. 
Following Eq.(\ref{EV2}) this is equivalent to demand
\begin{equation}
\epsilon(t) g(t)  = \frac{d a(t)}{dt} \label{track}
\end{equation}
again assuming $h(t) = 0$. This is usually called {\bf tracking} in the
literature. Tracking is hard to achieve without additional clever tricks
because Eq.(\ref{track}) is ill-defined whenever $g(t) = 0$ and 
$da(t)/dt \ne 0$,
and there is no (easy) way to know a priori (without previously writing
the field) what trajectories $a(t)$ are possible.

I didn't elaborate on what $\hat{A}$ is. In principle it can be any
operator. To get a feeling of Eq.(\ref{Elc}) I'll give you now two
examples.

\subsection{Slowing a charged particle}

So imagine that we have a free charged particle, $\hat{H}_0 = \hat{p}^2/2m$
and we want to slow it down with a field, 
$$ \frac{d \langle \hat{p}^2 \rangle }{dt} \le 0$$
What field should we used? First note that 
$$\left[ \hat{p}^2, \hat{H}_0 \right] = 0$$ 
and 
$$\left[ \hat{p}^2, \hat{\mu} \right] = \hat{p} \left[ \hat{p}, \hat{\mu} 
\right] + \left[ \hat{p}, \hat{\mu} \right] \hat{p} = 2i\hat{p}$$ 
(I am still using atomic units, where $\hat{\mu} = -z$).
Therefore, from Eq.(\ref{Elc}) we find
\begin{equation}
\epsilon_{lc}(t) = -2 f(t) \langle \psi(t) |  \hat{p} | \psi(t)\rangle
\end{equation}

This is just telling you that the field must oppose the momentum of the
electron, and the faster the electron goes, the stronger $\epsilon_{lc}(t)$
must be in order to stop the particle.

\subsection{Population transfer}

It is time to go back to our favorite operator, $\hat{P} = |\varphi_t\rangle
\langle \varphi_t |$ and see what LC tells us about how the field must
be in order to force a monotonic increase in $P_{0t}$ over time.
This is a nice exercise to compare the OC and LC procedures.

Notice that since I didn't write what $\hat{A}$ is in Eq.(\ref{Elc}), I was 
not very careful regarding the units of the different expressions, or put it 
differently, depending on what $\hat{A}$ was, $f(t)$ should have different 
units. It is time to be careful and let $f(t)$ have the (expected) dimensions
of a field, as we did with the optimal control equations.
Then following the same dimensional arguments, for $\hat{A} \equiv \hat{P}$
we should write Eq.(\ref{Elc}) as

\begin{equation}
\epsilon_{lc}(t) = -2 T f^2(t) 
\mathrm{Im} \left( \langle \psi(t) | \hat{P} \mu | \psi(t) \rangle \right)
\end{equation}
where I just changed $f(t) \rightarrow f^2(t)$ and added a $T$. 
This looks very much as Eq.(\ref{Eop}) except that $\langle \psi(t) | \hat{P}$
is not $\langle \chi(t) |$. Remember [Eq.(\ref{chit})] that
$\langle \chi(t) | = \langle \psi(T) | \hat{P} \hat{U}(T,t;\epsilon_{oc})$
If you want, you can call $\tilde{\chi}(t) = \hat{P} \psi(t)$
and $\tilde{\psi}(t) = \mu \psi(t)$. The fields are the overlap of these
two functions.
Whereas in LC we overlap $\tilde{\psi}(t)$ with the instantaneous projection
of $\psi(t)$ by $\hat{P}$, in OC we overlap $\tilde{\psi}(t)$ with the
projection of $\psi(T)$ at the future (at final time), which we then
propagate all the way down to the present. Let's re-write the equations
once more focusing on the bras (to turn up-side-down the expressions is the 
best way to learn what they mean, how they act). 
I will now use the fact that
$$\langle \chi(t) | = \langle \psi(T) | \hat{P} \hat{U}(T,t;\epsilon_{oc}) =
\langle\psi(t) | \hat{U}(t,T;\epsilon_{oc})\hat{P}\hat{U}(T,t;\epsilon_{oc})$$
In LC
\begin{equation}
\epsilon_{lc}(t) = -2 T f^2(t) 
\mathrm{Im} \left( \langle \psi(t) | \hat{P} | \tilde{\psi}(t) \rangle \right)
\end{equation}
In OC
\begin{equation}
\epsilon_{oc}(t) = -2 T f^2(t) 
\mathrm{Im} \left( \langle \psi(t) | \hat{U}(t,T;\epsilon_{oc}) \hat{P} 
\hat{U}(T,t;\epsilon_{oc}) | \tilde{\psi}(t) \rangle \right)
\end{equation}

The last equation is telling you that in OC the projection operator to your
target state is much more complex, it is taking into account the dynamics of 
the system. When 
$\hat{P}$ commutes with the time evolution operator 
$\hat{U}(T,t;\epsilon_{oc})$,
$$ \hat{U}(t,T;\epsilon_{oc}) \hat{P} \hat{U}(T,t;\epsilon_{oc}) = 
\hat{U}(t,T;\epsilon_{oc}) \hat{U}(T,t;\epsilon_{oc}) \hat{P} = \hat{P}$$
everything is much simpler.
Only then you should do the projection at each instant of time, without
care of the future dynamics, forcing the monotonic increase in the population 
transfer.
Only then the optimal field is just the local field.
Unfortunately it can be shown that this recipe only works when the dynamics
is trivial. The LC field is (almost) never optimal. You can (almost) always
improve its final yield with an optimized field version. But then LC is
much faster and computationally friendly.

\newpage

\lecture[]{2}{Quantum Optimal Control Algorithms}

\section{Quantum Optimal Control Algorithms}

You have seen that the instantaneous gradient of the functional with
respect to the field is
\begin{equation}
\frac{\delta J_t}{\delta\epsilon(t)} \equiv \frac{\delta J}{dt \delta\epsilon(t)} = -2 \left\{ \frac{1}{\alpha} \mathrm{Im} 
\left[ \langle \chi(t) | \mu |\psi(t)\rangle 
\right] + \frac{1}{Tf^2(t)} \left[ \epsilon(t) - \epsilon_\mathrm{ref}(t)
\right]  \right\}  \label{gradl2}
\end{equation}
where the function $\chi(t)$ obeys the same TDSE with final-time
boundary conditions and $\epsilon_\mathrm{ref}(t)$ is a reference
field with respect to which the penalty fluency is considered (so
we penalize deviations from this field)< $\alpha$ is a characteristic
value for $\langle \psi(T) | \hat{P} | \psi(T) \rangle$ and carries the units
of the positive operator $\hat{P}$. In most cases in these notes
$\epsilon_\mathrm{ref}(t)$ is set equal to zero and $\alpha = 1$ (as
when $\hat{P}$ is a projector operator with a natural desirable maximum
of $1$).

At the singular points the gradient is equal to zero defining the
optimal pulse. Thus, I showed that the optimal control field should obey 
the nonlinear (implicit) equation
\begin{equation}
\epsilon_\mathrm{OC}(t) = \epsilon_\mathrm{ref}(t) + \frac{T}{2} f^2(t) 
g_\mu(t) \label{Eopl2}
\end{equation}
where 
\begin{equation}
g_\mu(t) = -\frac{2}{\alpha} \mathrm{Im} \left[ \langle \chi(t) | \mu | 
\psi(t) \rangle \right] = 
\frac{\delta \langle \psi(T) | \hat{P} | \psi(T) \rangle}{\delta \epsilon(t)dt}  
\label{gmu}
\end{equation}
is the instantaneous gradient of the final target with respect 
to the field, and plays a most important role in quantum control.
In extended form, Eq(\ref{Eopl2}) can be written as:
\begin{equation}
\epsilon_\mathrm{OC}(t) = \epsilon_\mathrm{ref}(t) - \frac{T}{\alpha} f^2(t) 
\mathrm{Im}\left[\langle\psi(T)|\hat{P}\hat{U}\left(T,t;\epsilon_\mathrm{OC}(t)
\right) \mu \hat{U}\left(t,0;\epsilon_\mathrm{OC}(t)\right)| 
\psi_i \rangle \right] \label{Eope}
\end{equation}

Eq.(\ref{Eope}) [or (\ref{Eopl2})] can be solved only by iterative procedures.
Here we are going to study the most common algorithms that solve the 
optimal control equations and study their convergence properties.
Since the algorithms imply iterative procedures, I will use
$\epsilon^{(k)}(t)$ and $\epsilon^{(k+1)}(t)$ as two consecutive
steps in the long run to obtain good (optimal) solutions. To start the algorithm
running you always need an initial guess $k=0$. Whenever you have
a reference pulse, the choice $\epsilon^{(0)}(t) = \epsilon_\mathrm{ref}(t)$
is the most obvious. In some occasions a very
nonimaginative guess $\epsilon^{(0)}(t) = 0$ is sufficient to end-up
with good solutions. On the other hand you also need to give up at a
certain point. Normally this is achieved by imposing convergence conditions,
for instance 
$$  \int_0^T dt \frac{\left[ \epsilon^{(k+1)}(t) - 
\epsilon^{(k)}(t) \right]^2}{f(t)^2} \le T\kappa \hspace{1cm} \mathrm{or}
\hspace{1cm} J^{(k+1)} - J^{(k)} \le \kappa'$$
where $\kappa$ ($\kappa'$) are predetermined criteria for convergence.

In most of this section to save some space I will choose 
$\epsilon_\mathrm{ref}(t) = 0$. It can be recovered in a straightforward
way from all the equations by just changing the "zero" value of the field.
To simplify the notation and use Eq.(\ref{Eope}) I will call
$\psi^{(k)}(t)$ to the wave function obtained by propagating $\psi_i$ with
$\hat{U}\left(t,0;\epsilon^{(k)}(t)\right)$ and likewise
$\chi^{(k)}(t)$ to the performance function obtained by
propagating $\hat{P}\psi(T)$ with $\hat{U}\left(T,t;\epsilon^{(k)}(t)\right)$.

I will present now five algorithms of choice:
\begin{enumerate}[i.]
\item the gradient method~\cite{ShiJCP90,SolaJPCA98b}
\item the Krotov method~\cite{Tannor1992,SomloiCP93,PalaoPRA03}
\item the nested-Krotov method~\cite{ZhuJCP98,MadayJCP03,OhtsukiJCP04,OhtsukiPRA07,CastroPRE09}
\item Zhu-Rabitz scheme~\cite{ZhuJCP98b}
\item the two-point boundary-value quantum control paradigm (TBQCP)~\cite{HoPRE10,LiaoPRA11,HoCPC11}
\end{enumerate}

But first, let me tell you what does not work. Unfortunately,
from Eq.(\ref{Eope}) the obvious fixed-point iteration (or Newton method):
$$ \epsilon^{(k+1)}(t)= -\frac{T}{\alpha} f^2(t) \mathrm{Im} 
\left[ \langle\psi(T) | 
\hat{P} \hat{U}\left(T,t;\epsilon^{(k)}(t)\right) \mu 
 \hat{U}\left(t,0;\epsilon^{(k)}(t)\right)| \psi_i \rangle \right]$$
does not work in general.

\subsection{The gradient method}

The gradient method is based on choosing the field on
step $k+1$ as the field on step $k$ plus a contribution along the 
(approximate) gradient obtained with the same $k$ field. Thus
\begin{equation}
\frac{\delta J^{(k)}}{\delta\epsilon(t)dt} = 
\left[ g_\mu^{(k)}(t) - \frac{1}{Tf^2(t)} 
\epsilon^{(k)}(t) \right]  \label{grad2l2} 
\end{equation}
Then
\begin{equation}
\epsilon^{(k+1)}(t) = \epsilon^{(k)}(t) + \beta 
\frac{\delta J^{(k)}}{\delta\epsilon(t)dt}  \label{gradient}
\end{equation}
Where $\beta$ (a small but unknown positive number) must be found by linear
search.

It is not difficult to prove that by this choice $J^{(k+1)} \ge J^{(k)}$
(at least for a certain range of $\beta$). The argument is as follows.
Since, by integrating Eq.(\ref{gmu}) we find that
\begin{equation}
\Delta\left(\langle \psi^{(k)}(T)|\hat{P}|\psi^{(k)}(T)\rangle\right) =
\int_0^T g^{(k)}_\mu(t) \delta\epsilon^{(k)}(t) dt  \label{incgrad}
\end{equation}
taking into account only first order variations in the field, 
$\epsilon^{(k+1)}(t)^2 - \epsilon^{(k)}(t)^2 \approx 2\epsilon^{(k)}(t) 
\delta\epsilon^{(k)}(t)$,
\begin{equation}
\Delta J \approx \int_0^T \left[ g^{(k)}_\mu(t) - \frac{2}{Tf^2(t)} 
\epsilon^{(k)}(t) \right] \delta\epsilon^{(k)}(t) dt  \label{DJgrad}
\end{equation}
By making 
\begin{equation}
\epsilon^{(k+1)}(t) - \epsilon^{(k)}(t) \approx \delta\epsilon^{(k)}(t)
= \beta \left[ g^{(k)}_\mu(t) - \frac{2}{Tf^2(t)} \epsilon^{(k)}(t) \right]
\end{equation}
one is guaranteed to increase the functional.
This was our choice for $\delta\epsilon(t)$ [see Eq.(\ref{gradient})] and
indeed what I have shown is just a proof or test that Eq.(\ref{grad2l2}) 
[or Eq.(\ref{gradl2})] is the gradient of the functional with respect 
to the field. For the same reason, you can expect similar success if you 
play with more sophisticated gradient-line searches, like the conjugated 
gradient.
The larger the gradient or the larger $\beta$ are, the bigger will $\Delta J$ be.
However, if $\beta$ is too large, $J$ will no longer increase along the
gradient, which just provides the first order variation. 

How far can we move along the gradient, that is, what value should $\beta$
have, can only be tested numerically.
We need to create a sequence of $\beta_n$ and stop the search whenever
$J^{(k+1)}_{\beta_{n+1}} - J^{(k+1)}_{\beta_n} < 0$, after which we go
on for the next iteration, calculating a new gradient. Although more 
sophisticated line search exists, any such method is rather cumbersome.
It would be great to have a way to know $\beta$. 
Now suppose we had kept second order variations in the field but
{\em still assume} that we can use the instantaneous gradient of the final
target $g_\mu^{(k)}(t)$. We need to add the term depending on 
$\delta\epsilon^{(k)}(t)^2$ to Eq.(\ref{DJgrad}).
Then we get, after substitution of $\delta\epsilon(t)$ by its value
$$\Delta J \approx \int_0^T \left\{\beta\left[ g^{(k)}_\mu(t)-\frac{2}{Tf^2(t)} 
\epsilon^{(k)}(t) \right]^2 - \beta^2 \frac{1}{T f^2(t)} \left[ g^{(k)}_\mu(t) 
-\frac{2}{Tf^2(t)}\epsilon^{(k)}(t) \right]^2 \right\} dt $$  
\begin{equation}
= \int_0^T \left( \beta - \frac{1}{T f^2(t)} \beta^2 \right)
\left[ g^{(k)}_\mu(t) -\frac{2}{Tf^2(t)}\epsilon^{(k)}(t) \right]^2 dt
\hspace{4cm}
\end{equation}
For a constant $f(t) = f_0$ one can find that $\beta = Tf_0^2 / 2$ 
maximizes the integrand. Therefore you can estimate $\beta$.
However, once you go beyond first order, $g_\mu^{(k)}$ is no longer
the best function characterizing the sensitivity of the wanted dynamics to 
the effect of the field, but a very small modification does the job.
This is what the Krotov method gives you.

\subsection{The Krotov method}

When you study the convergence properties of algorithms, instead of $\delta J$ 
it is better to concentrate on the actual change of functional between
two consecutive steps,  $\Delta J = J^{(k+1)} - J^{(k)}$.
If the algorithms assures monotonic increase, $\Delta J$ should 
always be positive. In terms of the changes in the final state 
$\Delta\psi(T) = \psi^{(k+1)}(T) - \psi^{(k)}(T)$, and in the field,
$\Delta\epsilon(t) = \epsilon^{(k+1)}(t) - \epsilon^{(k)}(t)$,
$$\Delta J = \frac{1}{\alpha} \left[ \langle \psi^{(k+1)}(T) | \hat{P} | 
\psi^{(k+1)}(T) \rangle - \langle \psi^{(k)}(T) | \hat{P} | \psi^{(k)}(T) 
\rangle \right] - \frac{1}{T} \int_0^T dt \frac{\epsilon^{(k+1)}(t)^2 -
\epsilon^{(k)}(t)^2}{f^2(t)} $$ 
\begin{equation}
\;\;\;\;\;\;\;\; = 
\frac{1}{\alpha} \langle\Delta\psi(T) |\hat{P}| \Delta\psi(T)\rangle
+ \frac{2}{\alpha} \mathrm{Re}\left[\langle \psi^{(k)}(T)|\hat{P}| 
\Delta\psi(T)\rangle\right] -  \frac{1}{T} \int_0^T dt 
\frac{\Delta\epsilon(t)^2 + 2\epsilon^{(k)}(t) \Delta\epsilon(t)}{f^2(t)}
\label{DJ}
\end{equation}
Since the first term is always positive, you don't have to care about it.
$\Delta\psi(T)$ can be written in terms of $\Delta\epsilon(t)$ in the
same way than $\delta\psi(T)$ was written in terms of $\delta\epsilon(t)$
[Eq.(~\ref{dpsi})], but now we cannot drop things that depend on second
variations,
\begin{equation}
\Delta\dot{\psi} = -i \hat{H}_0 \Delta\psi + i\mu \epsilon^{(k+1)}
\psi^{(k+1)} - i \mu \epsilon^{(k)}\psi^{(k)} = 
-i \left[ \hat{H}_0 - \mu \epsilon^{(k)}\right] \Delta\psi + i\mu
\left[ \epsilon^{(k+1)} - \epsilon^{(k)} \right] \psi^{(k+1)}
\label{Ddotpsi}
\end{equation}
The inhomogeneous term now depends on the wave function $\psi^{(k+1)}$,
while the homogeneous term depends on the field at iteration $k$, 
$\epsilon^{(k)}(t)$. In the variational procedure we dropped the term 
$\Delta\epsilon(t)\Delta\psi(t)$ included in $\Delta\epsilon(t)\psi^{(k+1)}$, 
so that everything belonged to the same $k$ and only one field was invoked.
Now we have to be more careful. Then the change in the wave function at final
time between consecutive iterations is
\begin{equation}
\Delta\psi(T) =   i \int_0^T \hat{U}(T,t;\epsilon^{(k)}(t)) 
 \mu \Delta\epsilon(t) \psi^{(k+1)}(t) dt  \label{DpsiT}
\end{equation}
Introducing Eq.(\ref{DpsiT}) in Eq.(\ref{DJ}) we obtain, for the second term 
$$\frac{2}{\alpha} \mathrm{Re}\left[\langle \psi^{(k)}(T)|\hat{P}| 
\Delta\psi(T)\rangle\right] = 
\frac{2}{\alpha} \mathrm{Re}\left[ i \int_0^T \langle \psi^{(k)}(T)|\hat{P}
\hat{U}(T,t;\epsilon^{(k)}\mu| \psi^{(k+1)}(t) \rangle \Delta\epsilon(t) dt
\right]$$
\begin{equation} 
= -\frac{2}{\alpha} \int_0^T \mathrm{Im} \left[\langle \chi^{(k)}(t) |\mu| 
\psi^{(k+1)}(t) \rangle \right] \Delta\epsilon(t) dt = 
\int_0^T \tilde{g}^{(k,k+1)}_\mu(t) \Delta\epsilon(t) dt
\end{equation}
In order to distinguish the instantaneous gradient with this new modified or
generalized gradient function that is a function of two fields, we have added 
a tilde and two superindexes (only when the method converges 
$\epsilon^{(k+1)}(t) \approx \epsilon^{(k)}(t)$ the modified gradient becomes
the true gradient)
\begin{equation}
\tilde{g}^{(k,k+1)}_\mu(t) = -\frac{2}{\alpha}\mathrm{Im} \left[\langle 
\chi^{(k)}(t) |\mu| \psi^{(k+1)}(t) \rangle \right]  \label{gmuc}
\end{equation}
So we need to make the integrand
\begin{equation}
{\cal I} = \tilde{g}^{(k,k+1)}_\mu(t) \Delta\epsilon(t) 
- \frac{\Delta\epsilon(t)^2 + 2\epsilon^{(k)}(t) \Delta\epsilon(t)}{f^2(t) T}
\label{integrand}
\end{equation}
always positive.

There are many ways of making ${\cal I}$ positive. The Krotov method is based 
on maximizing it 
with respect to the variation $\Delta\epsilon(t)$
\begin{equation}
\frac{d{\cal I}(t)}{d\Delta\epsilon(t)} = \tilde{g}^{(k,k+1)}_\mu(t) 
- \frac{2 \Delta\epsilon(t) + 2\epsilon^{(k)}(t)}{f^2(t) T} = 0
\end{equation}
so that
\begin{equation}
\Delta\epsilon(t) = -\epsilon^{(k)}(t) + \frac{T}{2}f^2(t) 
\tilde{g}^{(k,k+1)}_\mu(t)
\end{equation}

The Krotov-optimized field for the step $k+1$ is therefore
\begin{equation}
\epsilon^{(k+1)}(t) = \epsilon^{(k)}(t) + \Delta\epsilon(t)
= - f^2(t) \frac{T}{\alpha} \mathrm{Im} \left[ \langle \chi^{(k)}(t) | 
 \mu | \psi^{(k+1)}(t) \rangle \right]  \label{EopKl2}
\end{equation}

This equation is very similar to Eq.(\ref{Eopl2}), but notice that while
$\chi(t)$ is propagated using the field obtained at the previous step
(the only one that is known at all times), $\psi(t)$ is propagated
with the new field. This implies immediate feedback for $\psi(t)$. 
Obviously some numerical approximation must be done at some point, 
since we do not know $\epsilon^{(k+1)}(t)$. 
However, numerically we can force $\epsilon^{(k+1)}(0)
= \epsilon^{(k)}(0)$ and then use this value, together with $\chi^{(k)}(t)$
to calculate via Eq.(\ref{EopKl2}) $\epsilon^{(k+1)}(\Delta t)$ and so on.
The immediate feedback will work better the smaller $\Delta t$ is.

To analyze the convergence properties of the Krotov method 
we just need to introduce the expression for the field in the respective
equation for the functional, Eq.(\ref{DJ}),
\begin{equation}
\Delta J = \frac{1}{\alpha} \langle\Delta\psi(T) |\hat{P}| \Delta\psi(T)\rangle
+ {\cal I}
\end{equation}
where the integrand is
\begin{equation}
{\cal I} =  \frac{Tf^2}{4} \left(\tilde{g}^{(k,k+1)}_\mu\right)^2 +
\frac{\left(\epsilon^{(k)}\right)^2}{Tf^2} > 0
\end{equation}

Therefore, Krotov guarantees a monotonic increase in the functional,
$\Delta J = J^{(k+1)} - J^{(k)} > 0$, and {\em we do not need any line search!}.
It does so providing you the {\em optimized step} (which is actually again of
the form $T f^2(t)/2$) along a {\em slightly changed direction} 
($\tilde{g}_\mu$ instead of $g_\mu$) and {\em erasing the previous version of 
the search} ($-\epsilon^{(k)}$). Krotov is not a conservative scheme at all.
By looking at the equations you can see that the term depending on 
$\tilde{g}^{(k,k+1)}_\mu(t)$ in $\Delta\epsilon(t)$ is the one responsible
for the increase in 
$\mathrm{Re}\left[\langle \Delta\psi(T) |\hat{P}| \Delta\psi(T)\rangle\right]$,
while the $-\epsilon^{(k)}(t)$ is mostly to avoid increasing the fluence
of the field, so you kind of remove all the components not along the
$k$-iteration-found modified gradient.
However, you can make smaller amendments in the control search if you
keep a reference field $\epsilon_\mathrm{ref}(t)$ so that you penalize
changes in the fluence with respect to this field. You can even actualize
the reference field at each iteration, 
$\epsilon_\mathrm{ref}(t) = \epsilon^{(k)}(t)$,
although this amounts to putting a strong brake in the control search that
Krotov explores.

Beware that maximizing ${\cal I}$, as Krotov does, does not imply 
maximizing $\Delta J$, since 
\begin{equation}
\langle\Delta\psi(T) |\hat{P}| \Delta\psi(T)\rangle = 
\int_0^T \!dt' \int_0^T \!dt'' \langle \psi^{(k+1)}(t') | \mu 
\hat{U}(t',T;\epsilon^{(k)}) \hat{P} \hat{U}(T,t'';\epsilon^{(k)}) \mu
\psi^{(k+1)}(t'') \rangle \Delta\epsilon(t')\Delta\epsilon(t'') 
\end{equation}
depends on the field on a complex way.
For instance, it is easy to check that ${\cal I}$ remains positive or zero 
for all values of $\Delta\epsilon(t)$ between $0$ and 
$Tf^2(t) g_\mu^{(k,k+1)}(t) - \epsilon^{(k)}(t)$, at which ${\cal I} = 0$. 
However, the first choice will obviously make $\Delta J = 0$ while the 
second will not.


\subsection{The nested Krotov method}


Notice that the last equality in Eq.(\ref{Ddotpsi}) is not unique. 
I could as well have written
\begin{equation}
\Delta\dot{\psi} = -i \hat{H}_0 \Delta\psi + i\mu \epsilon^{(k+1)}
\psi^{(k+1)} - i \mu \epsilon^{(k)}\psi^{(k)} = 
-i \left[ \hat{H}_0 - \mu \epsilon^{(k+1)}\right] \Delta\psi + i\mu
\Delta\epsilon \psi^{(k)}
\label{Ddotpsi2}
\end{equation}
where now the inhomogeneous term is $\psi^{(k)}$ while the propagator
in the homogeneous term depends on $\epsilon^{(k+1)}$.
Then the full variation of the wave function over the next iteration 
at final time would be written as
\begin{equation}
\Delta\psi(T) =   i \int_0^T \hat{U}(T,t;\epsilon^{(k+1)}(t)) 
 \mu \Delta\epsilon(t) \psi^{(k)}(t) dt  
\end{equation}
and the second term of $\Delta J$ in Eq.(\ref{DJ}) would be
\begin{equation}
\frac{2}{\alpha} \mathrm{Re}\left[\langle \psi^{(k)}(T)|\hat{P}| 
\Delta\psi(T)\rangle\right] 
= -\frac{2}{\alpha} \int_0^T \mathrm{Im} \left[\langle \chi^{(k+1)}(t) |\mu| 
\psi^{(k)}(t) \rangle \right] \Delta\epsilon(t) dt 
\end{equation}
By the obvious definition of
\begin{equation}
\tilde{g}^{(k+1,k)}_\mu(t) -\frac{2}{\alpha}\mathrm{Im} \left[\langle 
\chi^{(k+1)}(t) |\mu| \psi^{(k)}(t) \rangle \right] 
\end{equation}
maximizing the new integrand with respect to $\Delta\epsilon(t)$
we obtain the same Krotov equations. Therefore
\begin{equation}
\epsilon^{(k+1)}(t) = - f^2(t) \frac{T}{\alpha}
\mathrm{Im} \left[ \langle \chi^{(k+1)}(t) | \mu | 
\psi^{((k)}(t) \rangle \right]  \label{EopKm}
\end{equation}
is another way of implementing the Krotov scheme, where now
the performance function $\chi(t)$ gets immediate feedback whereas 
$\psi(t)$ experiences the usual iteration-to-iteration feedback.
There is no obvious advantage here from the previous Krotov expression, 
except when you have different functionals.
For instance, when $J$ includes time-dependent targets or time-dependent
penalties, one version of the scheme is easier to implement than the other.
The best choice depends on whether you want to maximize or minimize some
populations at intermediate times, as we shall see in the following section.

In the nested Krotov scheme one uses immediate feedback for both
the performance and the system's wave function.
Essentially this is done by dividing each iteration into a two-step
process, and making independent the fields that drive each of the
wave functions. So if we call $\epsilon_a^{(k)}(t)$ the auxiliary
field by which the dynamics of $\chi^{(k)}(t)$ is calculated, and
$\epsilon^{(k)}(t)$ to the field that optimizes the true dynamics,
that is, under which $\psi^{(k)}(t)$ is calculated, we can write
$$\epsilon_a^{(k+1)}(t) = - f^2(t) \frac{T}{\alpha} \mathrm{Im} 
\left[ \langle \chi^{(k+1)}(t) | \mu | \psi^{(k)}(t) \rangle \right] $$ 
and
\begin{equation}
\epsilon^{(k+1)}(t) = - f^2(t) \frac{T}{\alpha} \mathrm{Im} 
\left[ \langle \chi^{(k+1)}(t) | \mu | \psi^{(k+1)}(t) \rangle \right]
\end{equation}
or writing explicitly the TDSE
\begin{equation}
i\frac{\partial}{\partial t}\chi^{(k+1)}(t) = \hat{H}_0\chi(t) + 
\mu f^2(t) \frac{T}{\alpha} \mathrm{Im} \left[ \langle \chi^{(k+1)}(t) | \mu 
| \psi^{(k)}(t) \rangle \right] \chi^{(k+1)}(t) \label{1}
\end{equation}
\begin{equation}
i\frac{\partial}{\partial t}\psi^{(k+1)}(t) = \hat{H}_0\psi(t) + 
\mu f^2(t) \frac{T}{\alpha} \mathrm{Im} \left[ \langle \chi^{(k+1)}(t) | \mu 
| \psi^{(k+1)}(t) \rangle \right] \psi^{(k+1)}(t) \label{2}
\end{equation}
You need to solve in a previous iteration $\psi^{(k)}(t)$ and then
actualize immediately (from one time step to the next one) the value
of the field $\epsilon_a(t)$ that is used to propagate $\chi^{(k+1)}(t)$
from $t=T$ to $t=0$. Once you know $\chi^{(k+1)}(t)$, you can propagate 
$\psi^{(k+1)}(t)$ again actualizing the value of the new field 
$\epsilon^{(k+1})(t)$ at each time step.

It is not difficult to prove that the nested Krotov scheme makes
$\Delta J > 0$, but once you have two independent fields, you cannot
prove that your choices maximize ${\cal I}$.
For instance, you can generally write Eq.(\ref{Ddotpsi}) or Eq.(\ref{Ddotpsi2})
 in yet another form
\begin{equation}
\Delta\dot{\psi} = -i \left[ \hat{H}_0 \Delta\psi - \mu \epsilon_a^{(k)}\right]
\Delta\psi + i \mu \left(\epsilon^{(k+1)}-\epsilon_a^{(k)}\right) \psi^{(k+1)} 
+ i \mu \left(\epsilon_a^{(k)} -\epsilon^{(k)}\right) \psi^{(k)} 
\label{Ddotpsigen}
\end{equation}
so that
\begin{equation}
\Delta\psi(T) =   i \int_0^T \hat{U}(T,t;\epsilon_a^{(k)}) 
 \mu \Delta\epsilon_2(t) \psi^{(k+1)}(t) dt + i 
\int_0^T \hat{U}(T,t;\epsilon_a^{(k)}) \mu \Delta\epsilon_1(t) \psi^{(k)}(t) dt 
\end{equation}
where I defined $\Delta\epsilon_2 = \epsilon^{(k+1)}-\epsilon_a^{(k)}$
and $\Delta\epsilon_1 = \epsilon_a^{(k)} -\epsilon^{(k)}$.
$\epsilon_a^{(k)}$ is the field that will propagate $\chi^{(k)}(t)$, while 
$\epsilon^{(k)}$, $\epsilon^{(k+1)}$ propagate $\psi^{(k)}$ and $\psi^{(k+1)}$
respectively. Both of them appear in Eq.(\ref{Ddotpsigen}), requiring
two modified gradients
\begin{equation} 
\tilde{g}^{(k,k)}_{\mu} = -\frac{2}{\alpha}\mathrm{Im} \left[\langle 
\chi^{(k)}(t) |\mu| \psi^{(k)}(t) \rangle \right] 
\end{equation}
and
\begin{equation} 
\tilde{g}^{(k,k+1)}_{\mu} = -\frac{2}{\alpha}\mathrm{Im} \left[\langle 
\chi^{(k)}(t) |\mu| \psi^{(k+1)}(t) \rangle \right] 
\end{equation}
(Although $\tilde{g}^{(k,k)}_{\mu}$ has two equal indexes $k$ for both wave 
functions it is a two-pulse modified gradient and deserves the tilde 
because the field propagating $\chi$ 
is not the same field that propagates $\psi$.)
With the definition of the modified gradients working exactly as before
we obtain in $J$ an integrand of the form
\begin{equation}
{\cal I} = \tilde{g}^{(k,k+1)}_{\mu}(t) \Delta\epsilon_2(t) +
\tilde{g}^{(k,k)}_{\mu_1}(t) \Delta\epsilon_1(t)
- \frac{\Delta\epsilon_2(t)\left(\epsilon_a^{(k)}(t)+\Delta\epsilon_2(t)\right)
+ \Delta\epsilon_1(t)\left(\epsilon_a^{(k)}(t)-\Delta\epsilon_1(t)\right)}
{T f^2(t)}
\end{equation}
which has no maximum. 
But it can be proved that any choice of the type
(with a slight change of the index order, by which the initial guess field 
is already $k=1$, to allow generalization with simpler expressions),
\begin{equation}
\epsilon_a^{(k)}(t) = \left(1 -\eta\right) \epsilon^{(k)}(t) 
+ \eta f^2(t) \frac{T}{2} \tilde{g}^{(k,k)}_{\mu}(t)
\end{equation}
\begin{equation}
\epsilon^{(k+1)}(t) = \left(1 -\zeta\right) \epsilon_a^{(k)}(t)
+ \zeta f^2(t) \frac{T}{2} \tilde{g}^{(k,k+1)}_{\mu}(t)
\end{equation}
for any value $\eta, \zeta \in [0,2]$ (excluding $\eta = \zeta = 0$), 
implies monotonic increase in $\Delta J$. 
You can easily see that Krotov's corresponds to $\eta = 0$, $\zeta = 1$ 
or $\eta = 1$, $\zeta = 0$, while the nested-Krotov scheme implies 
$\eta = 1$ and $\zeta = 1$. Algorithms with parameters $[2,0]$ or $[0,2]$ 
make ${\cal I} = 0$. 
Then $\Delta J > 0$ comes uniquely from $\langle \Delta
\psi(T) |\hat{P}|\Delta\psi(T)\rangle > 0$. And indeed, from numerical
studies, $\zeta=2$ and $\eta = 0$ seems a particularly good choice.
From this general outset it seems remarkable that the simplest Newton
scheme allowed to at the beginning
\begin{equation}
\epsilon^{(k+1)}(t) = f^2(t) \frac{T}{2} g_\mu^{(k)}(t)
\end{equation}
performs so poorly.

\subsection{Zhu-Rabitz scheme}

Now we are going to present a different (though similiarly looking)
algorithm, that works for a particular class of positive semi-definite
operators, namely projectors $\hat{P} = |\psi_f\rangle \langle \psi_f|$.
This allows to write the control equations in a very symmetrical way
and exploit some interesting time-invariants.
To make the symmetry more apparent I write $\psi_1(t)$ to the wave function
propagated from the initial state $\psi_i$ at time $t=0$, and
$\psi_2(t)$ to the wave function propagated from the target state $\psi_f$
at time $t=T$.

Starting from the usual unconstrained functional (I make $\alpha = 1$)
\begin{equation}
J_{fi} = |\langle \psi_f |\psi_1(T)\rangle|^2 - \frac{1}{T} \int_0^T
\frac{\epsilon^2(t)}{f^2(t)} dt - 2 \mathrm{Re} \left[ \int_0^T \langle{\chi(t)
\left| \frac{\partial}{\partial t} + i\hat{H}(t) \right| \psi_1(t)}\rangle 
\right] dt 
\end{equation}
since $|\chi(T)\rangle = | \psi_f \rangle \langle \psi_f |\psi_1(T)\rangle$,
\begin{equation}
|\chi(t)\rangle = \hat{U}(t,T)|\chi(T)\rangle = 
\hat{U}(t,T)|\psi_f\rangle \langle \psi_f |\psi_1(T)\rangle =
|\psi_2(t)\rangle  \langle \psi_f |\psi_1(T)\rangle
\end{equation}
I can then change my performance function for $\psi_2(t)$, incorporating
$\langle \psi_f |\psi_1(T)\rangle$ to the unconstrained functional, making the
equations more symmetric,
\begin{equation}
J_{fi} = |\langle \psi_f |\psi_1(T)\rangle|^2 - \frac{1}{T} \int_0^T 
\frac{\epsilon^2(t)}{f^2(t)} dt - 2 \mathrm{Re} \left[ \langle\psi_1(T) |
\psi_f \rangle \int_0^T \langle{\psi_2(t)
\left| \frac{\partial}{\partial t} + i\hat{H}(t) \right| \psi_1(t)}\rangle 
\right] dt 
\end{equation}
for which
$$\epsilon_\mathrm{OC}(t) = -T f^2(t)\mathrm{Im}\left[\langle
\psi_1(T)|\psi_f\rangle\langle\psi_f | \hat{U}\left(T,t;\epsilon_\mathrm{OC}
\right) \mu \hat{U}\left(t,0;\epsilon_\mathrm{OC}\right) |\psi_i\rangle
\right]$$
\begin{equation}
= -T f^2(t)\mathrm{Im}\left[ \langle \psi_1(T)|\psi_f\rangle
\langle \psi_2(t) | \mu | \psi_1(t) \rangle \right] \hspace{2.5cm}
\label{EOCP}
\end{equation}
completed with the dynamical (TDSE) equations:
\begin{equation}
\frac{\partial}{\partial t}\psi_1(t) = -i\left[ \hat{H}_0 
-\mu\epsilon_\mathrm{OC}(t)\right] \psi_1(t), \hspace{0.5cm} \mathrm{with}
\;\; \psi_1(0) = \psi_i
\end{equation}
\begin{equation}
\frac{\partial}{\partial t}\psi_2(t) = -i\left[ \hat{H}_0 
-\mu\epsilon_\mathrm{OC}(t)\right] \psi_2(t), \hspace{0.5cm} \mathrm{with}
\;\; \psi_2(T) = \psi_f
\end{equation}
Because 
\begin{equation}
\frac{d}{dt} \langle \psi_1(t)|\psi_2(t)\rangle = \langle \dot{\psi}_1(t) 
|\psi_2(t)\rangle + \langle \psi_1(t)|\dot{\psi}_2(t)\rangle =
i\langle \psi_1(t) |\hat{H}(t) | \psi_2(t)\rangle 
-i \langle \psi_1(t) |\hat{H}(t) | \psi_2(t)\rangle = 0 
\end{equation}
we can find that 
\begin{equation}
\langle \psi_1(t)|\psi_2(t)\rangle = \langle \psi_1(T) |\psi_2(T) \rangle
= \langle \psi_1(T) |\psi_f\rangle = \langle \psi_i | \psi_2(0) \rangle
\end{equation}
is a time invariant of the dynamics.
This provides extra flexibility to the scheme.

Zhu and Rabitz propose a nested iteration scheme of the type
\begin{equation}
\epsilon_1^{(k+1)}(t) = -T f^2(t) \mathrm{Im} \left[ \langle \psi_1^{(k+1)}(t) |
\psi_2^{(k)}(t) \rangle \langle \psi_2^{(k)}(t) |\mu |\psi_1^{(k+1)}(t)\rangle
\right]
\end{equation}
\begin{equation}
\epsilon_2^{(k+1)}(t) = -T f^2(t) \mathrm{Im} \left[ \langle \psi_1^{(k+1)}(t) |
\psi_2^{(k+1)}(t) \rangle \langle \psi_2^{(k+1)}(t) |\mu |\psi_1^{(k+1)}(t)
\rangle \right]
\end{equation}
where $\epsilon_1^{(k)}(t)$ propagates $\psi_1^{(k)}(t)$ while
$\epsilon_2^{(k)}(t)$ propagates $\psi_2^{(k)}(t)$.
Notice the particular choice for the term $\langle \psi_1(T)|\psi_f\rangle$
that was in Eq.(\ref{EOCP}). Making use of the time invariance they choose
to pick the overlap at all the times. 
The factors in both equations
are slightly different owing to the different way in which the fields
are actualized by immediate feedback, and this gives non trivial differences
between the different possible implementations.
The previous algorithm incorporates immediate feedback and is prepared
to start the first iteration ($k=0$) from $\psi_2(T) = \psi_f$ with a guess 
field. For this reason $\epsilon_2(t)$ (which is the analog of $\epsilon_a(t)$
in the nested Krotov scheme) is actualized after $\epsilon_1(t)$.
Of course the algorithm is symmetric and could be done in the opposite
order. Using similar techniques as the ones I showed previously one can
find 
\begin{equation}
\Delta J = \int_0^T \left[ \frac{\left( \epsilon_1^{(k+1)} -
\epsilon_1^{(k)} \right)^2 + \left( \epsilon_2^{(k+1)} - \epsilon_1^{(k+1)} 
\right)}{T f^2(t)} \right] dt
\end{equation}
The scheme converges monotonically and quadratically with respect to 
the changes in the fields between consecutive iterations. This is similar
to the nested Krotov, but different than Krotov's. We can not guarantee
that the choice of fields maximizes ${\cal I}$.

An alternative (apparently faster to converge) algorithm, among many others, is
given by a different choice of the factors, given by the set of equations
\begin{equation}
\epsilon_1^{(k+1)}(t) = -T f^2(t) \mathrm{Im} \left[ \langle \psi_i |
\psi_2^{(k)}(0) \rangle \langle \psi_2^{(k)}(t) |\mu |\psi_1^{(k+1)}(t)\right]
\end{equation}
\begin{equation}
\epsilon_2^{(k+1)}(t) = -T f^2(t) \mathrm{Im} \left[ \langle \psi_1^{(k+1)}(T) |
\psi_f \rangle \langle \psi_2^{(k+1)}(t) |\mu |\psi_1^{(k+1)}(t)\right]
\end{equation}

\subsection{The TBQCP paradigm}

The latest and newest algorithm from the pool that I will show you
exploits another time-invariant to seek for both iteration-to-iteration
monotonic increase and ``local'' control features.

We will first introduce a new function with very interesting properties 
in the control problem.
Consider you have a reference field $\epsilon^{(r)}(t)$ (it could be a guess
field or a $k-1$ iteration-optimal field). We want to focus now on
\begin{equation}
{\cal P}^{(r)}(t) = \hat{U}\left(t,T;\epsilon^{(r)}\right) \hat{P} 
\hat{U}\left(T,t;\epsilon^{(r)}\right)
\end{equation}
Notice that this is not the Heisenberg picture form of the operator,
because the time-ordering inside the time evolution operator is the
opposite from what it should (from $t$ to a reference time, in this case 
the final time $T$, instead of the opposite in Heisenberg's case).

${\cal P}^{(r)}(t)$ is a dynamical invariant of the reference dynamics.
If we evaluate the average value of ${\cal P}^{(r)}(t)$ with the reference
wave function $\psi^{(r)}(t)$ we obtain
$$\langle \psi^{(r)}(t) |{\cal P}^{(r)}(t) |\psi^{(r)}(t) \rangle =
\langle \psi_i | \hat{U}\left(0,t;\epsilon^{(r)}\right) 
\hat{U}\left(t,T;\epsilon^{(r)}\right) \hat{P} 
\hat{U}\left(T,t;\epsilon^{(r)}\right)
\hat{U}\left(t,0;\epsilon^{(r)}\right) | \psi_i \rangle $$
\begin{equation}
\hspace{5cm}= \langle \psi_i | \hat{U}\left(0,T;\epsilon^{(r)}\right) \hat{P} 
\hat{U}\left(T,0;\epsilon^{(r)}\right) | \psi_i \rangle = 
\langle \psi^{(r)}(T) | \hat{P} | \psi^{(r)}(T) \rangle
\end{equation}
Since $t$ in the left hand does not show in the right hand of the previous 
equation, the result does not depend on the time at which we perform the 
average, 
\begin{equation}
\langle \psi^{(r)}(0) |{\cal P}^{(r)}(0) |\psi^{(r)}(0) \rangle =
\langle \psi^{(r)}(t) |{\cal P}^{(r)}(t) |\psi^{(r)}(t) \rangle =
\langle \psi^{(r)}(T) |{\cal P}^{(r)}(T) |\psi^{(r)}(T) \rangle =
\langle \psi^{(r)}(T) | \hat{P} | \psi^{(r)}(T) \rangle
\end{equation}
but remember that the initial state is the same regarless of the pulse, so 
\begin{equation}
\langle \psi^{(r)}(0) |{\cal P}^{(r)}(0) |\psi^{(r)}(0) \rangle =
\langle \psi^{(k)}(0) |{\cal P}^{(r)}(0) |\psi^{(k)}(0) \rangle =
\langle \psi^{(r)}(T) | \hat{P} | \psi^{(r)}(T) \rangle
\end{equation}

The central quantity in the TBQCP paradigm is ${\cal P}^{(r)}(t)$ averaged 
with respect to the wave function driven by an optimal pulse $\epsilon^{(k)}$,
because, as we will see, 
$\langle \psi^{(k)}(t) | {\cal P}^{(r)}(t) | \psi^{(k)}(t) \rangle$ can
be used to track the performance of the optimization and it is, in effect,
a functional of both the optimal pulse and the reference pulse.

For instance, if we evaluate the average of ${\cal P}^{(r)}(t)$ with the 
optimized wave function at final time $t = T$, $\psi^{(k)}(T)$,
\begin{equation}
\langle \psi^{(k)}(T) | {\cal P}^{(r)}(T) | \psi^{(k)}(T) \rangle = 
\langle \psi^{(k)}(T) | \hat{P} | \psi^{(k)}(T) \rangle
\end{equation}
Then
$$\int_0^T dt \frac{d}{dt} \langle \psi^{(k)}(t) |{\cal P}^{(r)}(t) 
|\psi^{(k)}(t) \rangle = \langle {\cal P}^{(r)}(T) \rangle - 
\langle {\cal P}^{(r)}(0) \rangle $$
$$ =
\langle \psi^{(k)}(T) |{\cal P}^{(r)}(T) |\psi^{(k)}(T) \rangle -
\langle \psi^{(k)}(0) |{\cal P}^{(r)}(0) |\psi^{(k)}(0) \rangle $$
\begin{equation}
\hspace{2.0cm} = \langle \psi^{(k)}(T) | \hat{P} | \psi^{(k)}(T) \rangle -
\langle \psi^{(r)}(T) | \hat{P} | \psi^{(r)}(T) \rangle \label{integral}
\end{equation}
tracks the improvement on the final yield obtained as we change the
field from the reference to the optimized $k$ iteration.

On the other hand, $\langle \psi^{(k)}(t) | {\cal P}^{(r)}(t) | \psi^{(k)}(t) \rangle$ 
obeys the following dynamical equation
$$ \frac{d}{dt} \langle \psi^{(k)}(t) |{\cal P}^{(r)}(t) |\psi^{(k)}(t) \rangle =
\langle \dot{\psi}^{(k)}(t) |{\cal P}^{(r)}(t) |\psi^{(k)}(t) \rangle +
\langle \psi^{(k)}(t) |{\cal P}^{(r)}(t) |\dot{\psi}^{(k)}(t) \rangle +
\langle \psi^{(k)}(t) |\dot{{\cal P}}^{(r)}(t) |\psi^{(k)}(t) \rangle $$
\begin{equation}
=- i \langle  \psi^{(k)}(t) | \left[ {\cal P}^{(r)}(t) , \hat{H}_0 
- \mu \epsilon^{(k)}(t) \right] | \psi^{(k)}(t) \rangle +
\langle \psi^{(k)}(t) |\dot{{\cal P}}^{(r)}(t) |\psi^{(k)}(t) \rangle
\end{equation}
Since
$$
\dot{{\cal P}}^{(r)}(t) = \dot{\hat{U}}\left(t,T;\epsilon^{(r)}\right) \hat{P} 
\hat{U}\left(T,t;\epsilon^{(r)}\right) + \hat{U}\left(t,T;\epsilon_r\right) \hat{P}
\dot{\hat{U}}\left(T,t;\epsilon^{(r)}\right) \hspace{2cm}$$
\begin{equation}
= i \left[ {\cal P}^{(r)}(t) , \hat{H}_0 - \mu \epsilon^{(r)}(t) \right] \hspace{5cm}
\end{equation}
then
\begin{equation}
\frac{d}{dt} \langle \psi^{(k)}(t) |{\cal P}^{(r)}(t) |\psi^{(k)}(t) \rangle =
i \langle  \psi^{(k)}(t) | \left[ {\cal P}^{(r)}(t) , \mu \right] | \psi^{(k)}(t) 
\rangle \left( \epsilon^{(k)}(t) - \epsilon^{(r)}(t) \right)
\end{equation}
Because of its importance, we give the commutator a special name, the
two-pulse overlap function
\begin{equation}
f^{(r,k)}_\mu(t) = \frac{i}{\alpha} \langle \psi^{(k)}(t) | \left[ {\cal P}^{(r)}(t) 
, \mu \right] | \psi^{(k)}(t) \rangle = 
- \frac{2}{\alpha} \mathrm{Im}\left( \langle \psi^{(k)}(t) | {\cal P}^{(r)}(t) \mu
 | \psi^{(k)}(t) \rangle \right)
\end{equation}
where as before, we add the $\alpha$ to keep track of the units of $\hat{P}$
($f^{(r,k)}_\mu(t)$ has units of inverse time and field; an $\hbar$ is always
missing in my equations).
The name recalls the two-pulse gradient function and this is on purpose, since 
$f^{(r,k)}_\mu(t)$ is very similar to the gradient function $g^{(k)}_\mu(t)$. 
Remember that
$$g^{(k)}_\mu(t) = -\frac{2}{\alpha} \mathrm{Im} \left( \langle \chi^{(k)}(t)
|\mu |\psi^{(k)}(t) \rangle \right)$$
Since
$$\chi^{(k)}(t) = \hat{U}\left(t,T;\epsilon^{(k)}\right) \hat{P} \psi^{(k)}(T)
=\hat{U}\left(t,T;\epsilon^{(k)}\right) \hat{P}\hat{U}\left(T,t;\epsilon^{(k)}
\right) \psi^{(k)}(t)$$
\begin{equation}
g^{(k)}_\mu(t) = -\frac{2}{\alpha} \mathrm{Im} \left( \langle \psi^{(k)}(t) | 
\hat{U}\left(t,T;\epsilon^{(k)}\right) \hat{P} \hat{U}
\left(T,t;\epsilon^{(k)}\right) \mu |\psi^{(k)}(t) \rangle \right)
\end{equation}
You can thus see that {\em the two-pulse gradient coincides with the 
gradient for the reference field}, $f_\mu^{(r,r)}(t) = g^{(r)}_\mu(t)$
(which is obviously equal to the two-pulse gradient using the same field
for both functions).

Suppose now that we use the previous iteration field as a reference. Is it
equivalent to the two-pulse modified gradient $\tilde{g}^{(k,k+1)}_\mu(t)$?
$$g^{(k,k+1)}_\mu(t) = -\frac{2}{\alpha} \mathrm{Im} \left( \langle
\psi^{(k)}(t) | \hat{U}\left(t,T;\epsilon^{(k)}\right) \hat{P} 
\hat{U} \left(T,t;\epsilon^{(k)}\right) \mu |\psi^{(k+1)}(t) \rangle \right) 
\hspace{4cm} $$
\begin{equation}
\ne -\frac{2}{\alpha} \mathrm{Im} \left( \langle \psi^{(k+1)}(t) | 
\hat{U}\left(t,T;\epsilon^{(k)}\right) \hat{P} 
\hat{U} \left(T,t;\epsilon^{(k)}\right) \mu |\psi^{(k+1)}(t) \rangle \right)
= f^{(k,k+1)}_\mu(t)
\end{equation}
Thus only on convergence the two-pulse function is the gradient and hence
also the two-pulse modified gradient.

Now inserting the dynamical equation for  
$\langle \psi^{(k)}(t) | {\cal P}^{(r)}(t) | \psi^{(k)}(t) \rangle$
on Eq.(\ref{integral}) we arrive to the main result of the TBQCP paradigm
\begin{equation}
\int_0^T f^{(r,k)}_\mu(t)\left( \epsilon^{(k)}(t) -\epsilon^{(r)}(t)\right) dt
= \frac{1}{\alpha}\left[ \langle \psi^{(k)}(T) | \hat{P} | \psi^{(k)}(T) 
\rangle - \langle \psi^{(r)}(T) | \hat{P} | \psi^{(r)}(T) \rangle \right]
\end{equation} 
An equation that clearly resembles Eq.(\ref{incgrad}) for the gradient.
Clearly, if we want to increase the performance of the reference field,
we just need to make 
$\epsilon^{(k)}(t) - \epsilon^{(r)}(t) \propto f^{(r,k)}_\mu(t)$.
Or more precisely
\begin{equation}
\epsilon^{(k)}(t) = \epsilon^{(r)}(t) + \eta S(t) \left( f^{(r,k)}_\mu(t) 
+ f_\perp(t) \right)
\end{equation}
where $S(t)$ is a positive envelope function, $\eta$ a positive constant 
and $f_\perp(t)$ is a function of time in the null space of $f^{(r,k)}_\mu(t)$
(the trivial being just zero), e.g.
\begin{equation}
\int_0^T S(t) f^{(r,k)}_\mu(t) f_\perp(t) dt = 0
\end{equation}

A typical implementation of the optimization scheme uses the recurrence
relation
\begin{equation}
\epsilon^{(k+1)}(t) = \epsilon^{(k)}(t) + \left( 1 - \zeta^{(k+1)}\right)
\eta S(t) f_\mu^{(k,k+1)}(t) + \eta S(t) f^{(k+1)}(t)
\end{equation}
with
\begin{equation}
f_\mu^{(k,k+1)}(t) = -\frac{2}{\alpha} \mathrm{Im}\left( \langle
\psi^{(k+1)}(t) | {\cal P}^{(k)}(t)\mu | \psi^{(k+1)}(t) \rangle\right)
\end{equation}
and
\begin{equation}
\zeta^{(k+1)} = \frac{\int_0^T S(t) f_\mu^{(k,k+1)}(t) f^{(k+1)}(t) dt}
{\int_0^T S(t) \left[ f_\mu^{(k,k+1)}(t) \right]^2 dt}
\approx \frac{\int_0^T S(t) f_\mu^{(k-1,k)}(t) f^{(k+1)}(t) dt}
{\int_0^T S(t) \left[ f_\mu^{(k-1,k)}(t) \right]^2 dt}
\end{equation}
where $f^{k}(t)$ can be any function, including zero. The approximation
in the last equation is needed as $\zeta$ can only be obtained at the
end of the iteration (it involves the integral over all time) and thus
cannot incorporate inmediate feedback.

For projector operators $\hat{P} = | \psi_f \rangle \langle \psi_f |$
a fast-kick-off monotonically convergent algorithm was proposed
\begin{equation}
\epsilon^{(k+1)}(t) = \epsilon^{(r)}(t) + \eta S(t) \epsilon_\gamma(t)
\end{equation}
with
\begin{equation}
\epsilon_\gamma(t) = \frac{f_\mu^{(k+1,r)}(t)}{\zeta_\gamma(t)} = 
-\frac{2}{\alpha} \mathrm{Im}\left( \frac{\langle \psi_1^{(k+1)}(t)|
\psi_2^{(r)}(t)\rangle}{\left| \langle \psi_1^{(k+1)}(t)| \psi_2^{(r)}(t)
\rangle \right|^\gamma} \times \langle \psi_2^{(r)}(t) |\mu|
\psi_1^{(k+1)}(t)\rangle \right)
\end{equation}
where $\psi_2^{(r)}(T) = \psi_f$ is only propagated by the (initially guessed) 
reference field, while $\psi^{(k+1)}_1(0) = \psi_i$ is propagated by 
$\epsilon^{(k+1)}(t)$ with immediate feedback. The other parameters
must be chosen as $0 \ge \gamma \ge 1$ and $\eta > 0$. 
In principle, the larger $\gamma$ and $\eta$ the faster the convergence.

\subsection{Similarities and differences between the schemes}

The variational condition of optimality requires penalizing the field
strength. Most control problems have dealt with maximizing positive operators
and penalizing field intensities. This 
square-in-$\psi(T)$-square-in-$\epsilon(t)$ functional has many mathematical
properties that favor robust solutions.

Using our nomenclature, optimality implies
\begin{equation}
\epsilon_\mathrm{OC}(t) = \epsilon_\mathrm{ref}(t) + \frac{T}{2}f^2(t) g_\mu(t)
\end{equation}
where $g_\mu(t)$ is the gradient of the final target.
%
Looking at the particular case of a projector operator the gradient
methods proposes the following recursive scheme
\begin{equation}
\epsilon^{(k+1)}(t) = \epsilon_\mathrm{ref}(t) + \left( 1 - 
\frac{\beta}{Tf^2(t)} \right) \epsilon^{(k)}(t) + \beta g_\mu^{(k)}(t) 
\end{equation}
with the suggestion, at least for constant and small $f$ that
$\beta \sim Tf^2/2$, so 
\begin{equation}
\epsilon^{(k+1)}(t) \sim \epsilon_\mathrm{ref}(t) + \frac{1}{2} 
\epsilon^{(k)}(t) + \frac{Tf^2}{2} g_\mu^{(k)}(t) 
\end{equation}
(From here you can stimate how small should the gradient be. For the
linear approximation to hold one can expect $T f^2 g_\mu^{(k)}(t) < 
\epsilon^{(k)}(t)$ and, assuming $f \sim \epsilon^{(k)}(t)$, that 
is necessary to have $g_\mu^{(k)}(t) < (T f)^{-1}$.)
The performance function $\chi^{(k)}(t)$ is always propagated backwards from
$\hat{P}\psi^{(k)}(T)$ by the same field than $\psi_i$,
$\epsilon_a^{(k)}(t) = \epsilon^{(k)}(t)$.

The Krotov scheme proposes
\begin{equation}
\epsilon^{(k+1)}(t) = \epsilon_\mathrm{ref}(t) + \frac{Tf^2}{2} 
\tilde{g}_\mu^{(k,k+1)}(t)  \label{Kro1}
\end{equation}
or
\begin{equation}
\epsilon^{(k+1)}(t) = \epsilon_\mathrm{ref}(t) + \frac{Tf^2}{2} 
\tilde{g}_\mu^{(k+1,k)}(t)  \label{Kro2}
\end{equation}
where the recursive relation must be understood as imposing 
$\epsilon_a^{(k+1)}(t) = \epsilon^{(k+1)}(t)$ at the end of the iteration
in order to return again to use Eq.(\ref{Kro1}) (or viceverse for the
second Krotov scheme, that is we actualize 
$\epsilon^{(k+1)}(t) = \epsilon_a^{(k+1)}(t)$ at the end of an iteration
to use again Eq.(\ref{Kro2})).
The most general version of the nested Krotov scheme requires
\begin{equation}
\epsilon^{(k+1)}(t) = \epsilon_\mathrm{ref}(t) + \left(1 -\zeta\right) 
\epsilon_a^{(k)}(t) + \zeta \frac{Tf^2(t)}{2} \tilde{g}_{\mu_2}^{(k,k+1)}(t)
\end{equation}
that must be completed with a different recursive relation for the iterative 
sequence of pulses that propagates $\chi^{(k)}(t)$, 
\begin{equation}
\epsilon_a^{(k)}(t) = \epsilon_\mathrm{ref}(t) + \left(1 -\eta\right) 
\epsilon^{(k)}(t) + \eta \frac{T f^2(t)}{2} \tilde{g}_{\mu_1}^{(k,k)}(t)
\end{equation}
The TBQCP in the simplest form (making $f_\perp(t) =0$) proposes
\begin{equation}
\epsilon^{(k)}(t) = \epsilon_\mathrm{ref}(t) + \eta S(t) 
f^{(\mathrm{ref},k)}_\mu(t) 
\end{equation}
For the projector $\hat{P} = |\psi_f\rangle\langle \psi_f|$,
 ${\cal P}^{(k)}(t) = |\psi_2^{(k)}(t)\rangle\langle\psi_2^{(k)}|$
and 
$$ -2 \mathrm{Im} \left[ \langle \psi_1^{(k+1)}(t) | \psi_2^{(k)}(t) \rangle 
\langle \psi_2^{(k)}(t) |\mu |\psi_1^{(k+1)}(t)\rangle \right] = 
f^{(k,k+1)}_\mu(t)$$
Therefore $\psi_i$ is propagated by $\epsilon(t)$ while $\psi_f$ [and not
$\chi(T)$] is propagated backwards by the reference field.
Using the TBQCP nomenclature, the Zhu-Rabitz scheme is equivalent to writing
\begin{equation}
\epsilon_1^{(k+1)}(t) = \epsilon_\mathrm{ref}(t)+\frac{T}{2} f_\mu^{(k,k+1)}(t)
\end{equation}
and
\begin{equation}
\epsilon_2^{(k+1)}(t) =\epsilon_\mathrm{ref}(t)+\frac{T}{2} f_\mu^{(k+1,k+1)}(t)
\end{equation}
where $\epsilon_1(t)$ propagates $\psi_i$ while $\epsilon_2(t)$ propagates
$\psi_f$ backwards.
Finally, the fast-kick-off monotonically convergent algorithm is
\begin{equation}
\epsilon^{(k+1)}_1(t) = \epsilon_\mathrm{ref}(t) + \eta S(t) 
\frac{f_\mu^{(\mathrm{ref},k+1)}(t)}{\zeta_\gamma(t)}
\end{equation}
with
\begin{equation}
\zeta_\gamma(t) = \left| \langle \psi_1^{(k+1)}(t)| \psi_2^{(\mathrm{ref})}(t)
\rangle \right|^\gamma
\end{equation}
where $\psi_2(t)$ is always propagated by $\epsilon_\mathrm{ref}(t)$.

\bibliographystyle{unsrt}
\bibliography{examples/references}

\end{document}